\documentclass[showpacs,amsmath,amssymb,aps,showkeys,floatfix,prd,a4paper]{revtex4}

\usepackage[dvips]{graphicx}
\usepackage{dcolumn}
\usepackage{bm}
\usepackage{epsfig}
\usepackage{amsfonts}
\usepackage{amssymb,amscd}

\def\lsim{\raise0.3ex\hbox{$<$\kern-0.75em\raise-1.1ex\hbox{$\sim$}}}
\def\gsim{\raise0.3ex\hbox{$>$\kern-0.75em\raise-1.1ex\hbox{$\sim$}}}

\def\beq{\begin{equation}}
\def\eeq{\end{equation}}
\def\bea{\begin{eqnarray}}
\def\eea{\end{eqnarray}}
\def\bq{\begin{quote}}
\def\eq{\end{quote}}

\newcommand{\rr}{\mbox{\boldmath $r$}}

\newcommand{\rb}{\mbox{\boldmath $b$}}

\newcommand{\rd}{\mbox{\boldmath $\Delta$}}

\def\gappeq{\mathrel{\rlap {\raise.5ex\hbox{$>$}}
{\lower.5ex\hbox{$\sim$}}}}

\def\lappeq{\mathrel{\rlap{\raise.5ex\hbox{$<$}}
{\lower.5ex\hbox{$\sim$}}}}

\def\Toprel#1\over#2{\mathrel{\mathop{#2}\limits^{#1}}}

\begin{document}


\title{Exclusive $\Upsilon$ photoproduction in hadronic collisions at CERN LHC energies }

\author{V.~P. Gon\c{c}alves $^1$, B. D. Moreira $^2$ and F. S. Navarra $^2$}
\affiliation{$^1$ High and Medium Energy Group, \\
Instituto de F\'{\i}sica e Matem\'atica, Universidade Federal de Pelotas\\
Caixa Postal 354, CEP 96010-900, Pelotas, RS, Brazil \\ $^2$ Instituto de F\'{\i}sica, Universidade de S\~{a}o Paulo,
C.P. 66318,  05315-970 S\~{a}o Paulo, SP, Brazil}

\date{\today}

\begin{abstract}

The exclusive $\Upsilon$ photoproduction   in proton-proton and  proton - nucleus  collisions at LHC energies is investigated using the color dipole formalism and considering different models for  the $\Upsilon$ wave function and  forward dipole - target scattering amplitude. Our goal is to update the color dipole predictions and  estimate the theoretical uncertainty present in these predictions. We present predictions for the kinematical ranges probed by the ALICE, CMS and LHCb Collaborations.
 
\end{abstract}

\maketitle

\section{Introduction}
\label{intro}

The experimental results from CDF  \cite{cdf} at Tevatron, STAR \cite{star} and PHENIX \cite{phenix} at RHIC and ALICE \cite{alice,alice2} and LHCb \cite{lhcb,lhcb2} at LHC for photon-induced processes 
 in hadronic collisions  have demonstrated in the last years that 
a detailed analysis is feasible and that the data can be used to constrain the description of the hadronic structure at high energies as well as to probe possible scenarios for the physics beyond the Standard Model 
 (For reviews see Ref. \cite{upc}). Recently, the status of    photon - photon and photon - hadron interactions in $pp/pA/AA$ collisions has been reviewed in a dedicated workshop at CERN \cite{workshop}. Moreover, the upcoming experimental data and new observables which could be studied in future runs of the LHC have been discussed in detail.  In particular, it is now clear that the first experimental data for the exclusive $\Upsilon$ photoproduction in $pp$ and $pPb$ collisions will be available in the next months. Such process was studied by several theoretical groups considering different formalisms and underlying assumptions 
\cite{vicmag_ups,frankfurt_ups,Schafer_ups,motyka_watt,cox_forshaw,Martin}. For example, the current predictions derived from the color dipole formalism \cite{vicmag_ups,motyka_watt,cox_forshaw} were obtained using different treatments for dependence of the cross section on the squared momentum transfer $t$ and distinct models for the $\Upsilon$ wave function  and/or for the forward dipole - proton scattering amplitude $\cal{N}$. Such differences render the interpretation of the results a hard task. 
Our goal in this paper is to update the color dipole predictions for the $\Upsilon$ production and compare the results obtained considering different models for $\cal{N}$ 
and for the $\Upsilon$  wave function and also different assumptions for the $t$-dependence of the cross section. 
We want to estimate the theoretical uncertainty present in the current predictions in the literature (For similar studies for the $J/\Psi$ and $\rho$ production see Refs. \cite{bruno,glauber_rho}, respectively). We start our study discussing the exclusive $\Upsilon$ photoproduction at HERA and compare our predictions with the scarce experimental data. After that we present our predictions for the rapidity distribution and total cross sections for the exclusive photoproduction of $\Upsilon$ in $pp$ collisions at $\sqrt{s} = 7,\,8$ and 14 TeV as well as in $pPb$ collisions at $\sqrt{s} = 5$ TeV. We also present our predictions of the total cross sections in the kinematical 
range probed by the LHCb Collaboration.

The paper is organized as follows. In the next section we present a brief review of photon - hadron 
interactions in $pp$ and $pPb$ collisions, as well as of the color dipole formalism for the exclusive $\Upsilon$ photoproduction. We also present the models for the dipole - target scattering 
amplitude and $\Upsilon$ wave functions used in our calculations. In Section \ref{resultados} we present our predictions 
for the exclusive photoproduction of $\Upsilon$ in $\gamma p/pp/pPb$ collisions and a comparison 
with the HERA data is also  shown. Finally, in Section \ref{sumario} we summarize our main conclusions.

\section{Exclusive $\Upsilon$ photoproduction in hadronic collisions}
\label{formulas}

In hadronic collisions at large impact parameter 
($b > R_{h_1} + R_{h_2}$) and at ultra relativistic energies the 
electromagnetic interaction  is expected  to be dominant. In this regime, 
the cross sections for a given process can be factorized in terms of the equivalent flux of 
photons of the hadron projectile and  the  photon-target production cross 
section \cite{upc}. In particular, the rapidity distribution for the exclusive $\Upsilon$ photoproduction  
in hadronic collisions is given by
\begin{eqnarray}
\frac{d\sigma \,\left[h_1 + h_2 \rightarrow   h_1 \otimes \Upsilon \otimes h_2\right]}{dY} = \left[\omega \frac{dN}{d\omega}|_{h_1}\,\sigma_{\gamma h_2 \rightarrow \Upsilon \otimes h_2}\left(\omega \right)\right]_{\omega_L} + \left[\omega \frac{dN}{d\omega}|_{h_2}\,\sigma_{\gamma h_1 \rightarrow \Upsilon \otimes h_1}\left(\omega \right)\right]_{\omega_R}\,
\label{dsigdy}
\end{eqnarray}
where the rapidity ($Y$) of the $\Upsilon$ in the final state is determined by the photon energy $\omega$ in the collider frame and by mass $M_{\Upsilon}$ of the vector meson [$Y\propto \ln \, ( \omega/M_{\Upsilon})$].
The symbol
$\otimes$ represents the presence of a rapidity gap in the final state and $\omega_L \, (\propto e^{-Y})$ and $\omega_R \, (\propto e^{Y})$ denote photons from the $h_1$ and $h_2$ hadrons, respectively.  
The  equivalent photon 
spectrum $\frac{dN}{d\omega}$ of a relativistic proton is given by  \cite{Dress},
\begin{eqnarray}
\frac{dN_{\gamma/p}(\omega)}{d\omega} =  \frac{\alpha_{\mathrm{em}}}{2 \pi\, \omega} \left[ 1 + \left(1 -
\frac{2\,\omega}{\sqrt{s_{NN}}}\right)^2 \right] 
\left( \ln{\Omega} - \frac{11}{6} + \frac{3}{\Omega}  - \frac{3}{2 \,\Omega^2} + \frac{1}{3 \,\Omega^3} \right) \,,
\label{eq:photon_spectrum}
\end{eqnarray}
with the notation $\Omega = 1 + [\,(0.71 \,\mathrm{GeV}^2)/Q_{\mathrm{min}}^2\,]$, 
$Q_{\mathrm{min}}^2= \omega^2/[\,\gamma_L^2 \,(1-2\,\omega /\sqrt{s_{NN}})\,] \approx 
(\omega/
\gamma_L)^2$, $\gamma_L$ is the Lorentz boost  of a 
single beam and $\sqrt{s_{\mathrm{NN}}}$ is 
 the c.m.s energy of the
hadron-hadron system. The equivalent photon flux of a nuclei is assumed to be given by \cite{upc}
\begin{eqnarray}
\frac{dN_{\gamma/A}\,(\omega)}{d\omega}= \frac{2\,Z^2\alpha_{em}}{\pi\,\omega}\, \left[\bar{\eta}\,K_0\,(\bar{\eta})\, K_1\,(\bar{\eta})+ \frac{\bar{\eta}^2}{2}\,{\cal{U}}(\bar{\eta}) \right]\,
\label{fluxint}
\end{eqnarray}
where   $\bar{\eta}=\omega\,(R_{h_1} + R_{h_2})/\gamma_L$, $K_{0,1}$ are the modified Bessel functions of second kind and  
${\cal{U}}(\bar{\eta}) = K_1^2\,(\bar{\eta})-  K_0^2\,(\bar{\eta})$, which is enhanced by a factor $Z^2$ in comparison to the proton one. It is important to observe that the photon fluxes, Eqs. (\ref{eq:photon_spectrum}) and (\ref{fluxint}), have support at small values of $\omega$, decreasing exponentially at large $\omega$. Consequently, the first term on the right-hand side of the Eq. (\ref{dsigdy}) peaks at positive rapidities while the second term peaks at negative rapidities. 
Moreover, given the photon flux, the study of the rapidity distribution can be used to constrain  the photoproduction cross section  at  a given energy. 
Finally, due to the differences between the equivalent photon flux of the proton and of the nucleus, the rapidity distribution of the $\Upsilon$'s produced in $pPb$ 
collisions will be asymmetric and determined by $\gamma p$ interactions, with the photon coming from the nucleus. In contrast, the rapidity distribution for $pp$ 
collisions will be symmetric with respect to $Y=0$. 

In the color dipole formalism the $\gamma h$  scattering is described  in the dipole frame, in which most of the energy 
is carried by the hadron, while the  photon  has
just enough energy to dissociate into a quark-antiquark pair
before the scattering. In this representation the probing
projectile fluctuates into a
quark-antiquark pair (a dipole) with transverse separation
$\rr$ long before the interaction, which then
scatters off the hadron \cite{nik}.
In this formalism, the 
scattering  
 amplitude for the diffractive photoproduction of an exclusive final state, 
such as a $\Upsilon$, in a $\gamma p$ collision is given by 
(See e.g. Refs. \cite{nik,vicmag_mesons,KMW,armesto_amir})
\begin{eqnarray}
 {\cal A}^{\gamma p \rightarrow \Upsilon p}(x,\Delta)  =  i
\int dz \, d^2\rr \, d^2\rb  e^{-i[\rb-(1-z)\rr].\rd} 
 \,\, (\Psi_{\Upsilon}^* \Psi) \,\,2 {\cal{N}}_p(x,\rr,\rb)
\label{sigmatot2}
\end{eqnarray}
where $(\Psi_{\Upsilon}^* \Psi)$ denotes the overlap of the photon and $\Upsilon$   
transverse wave functions. The variable  $z$ $(1-z)$ is the
longitudinal momentum fraction of the quark (antiquark),  $\Delta$ denotes the transverse 
momentum lost by the outgoing proton ($t = - \Delta^2$) and $x$ is the Bjorken variable. 
The variable $\rb$ is the transverse distance from the center of the target to the center of mass of the 
$q \bar{q}$  dipole and the factor  in the exponential  arises when one takes into account 
non-forward corrections to the wave functions \cite{non}.
 Moreover, ${\cal{N}}_p (x,\rr,\rb)$ denotes the non-forward scattering  amplitude of a dipole of size $\rr$ on the proton, which is  directly related to  the QCD 
dynamics (see below).
The differential cross section  for  exclusive $\Upsilon$ photoproduction is given by
\begin{eqnarray}
\frac{d\sigma}{dt} (\gamma p \rightarrow  \Upsilon p) = \frac{1}{16\pi} |{\cal{A}}^{\gamma p \rightarrow \Upsilon p}(x,\Delta)|^2\,(1 + \beta^2)\,R_g^2\,,
\label{totalcs}
\end{eqnarray}
where $\beta$ is the ratio of real to imaginary parts of the scattering
amplitude and  $R_g$ is the skewness factor, which is associated to the fact that the gluons attached to the $q\bar{q}$ pair can carry different light-cone fractions $x$, $x^{\prime}$ of the proton. In the limit that $x^{\prime} \ll x \ll 1$ and at small $t$ and assuming that the gluon density has a power-law form $xg \propto x^{-\lambda_e}$, it is given by \cite{Shuvaev:1999ce}
\begin{eqnarray}
\label{eq:Rg}
  R_g(\lambda_e) = \frac{2^{2\lambda_e+3}}{\sqrt{\pi}}
\frac{\Gamma(\lambda_e+5/2)}{\Gamma(\lambda_e+4)}, 
  \quad\text{with} \quad \lambda_e \equiv 
\frac{\partial\ln\left[\mathcal{A}(x,\,\Delta)\right]}{\partial\ln(1/x)}.
\end{eqnarray}
Moreover, $\beta$ can be calculated using dispersion relations, being given by
 $Re {\cal A}/Im {\cal A}=\mathrm{tan}\,(\pi \lambda_e/2)$. 
 The total cross section is given by 
\begin{eqnarray}
\sigma (\gamma p \rightarrow \Upsilon p) = \int_{-\infty}^0 \frac{d\sigma}{dt}\, dt \,\,.
\label{sctotal_intt}
\end{eqnarray}
In what follows, we will also calculate the total cross section considering an approximation frequently used in the literature, in which  an exponential Ansatz for the $t$-dependence is assumed for the differential cross section, which implies that
\begin{eqnarray}
\sigma (\gamma p \rightarrow \Upsilon p) =  \frac{1}{B_V}\,\left. \frac{d\sigma}{dt}\right|_{t=0} \,
\label{dsdt_approx}
\end{eqnarray}
where $B_{\Upsilon}$ is the slope parameter. As in Ref. \cite{cox_forshaw}, we will use in our calculations the following parametrisation
\begin{eqnarray}
B_{\Upsilon} = N\,\left[ \frac{14}{(M_{\Upsilon}/GeV)^{0.4}}+1  \right]                
\end{eqnarray}                                       
with $N= 0.55$ GeV$^{-2}$. 

\begin{figure}
\includegraphics[scale=0.35]{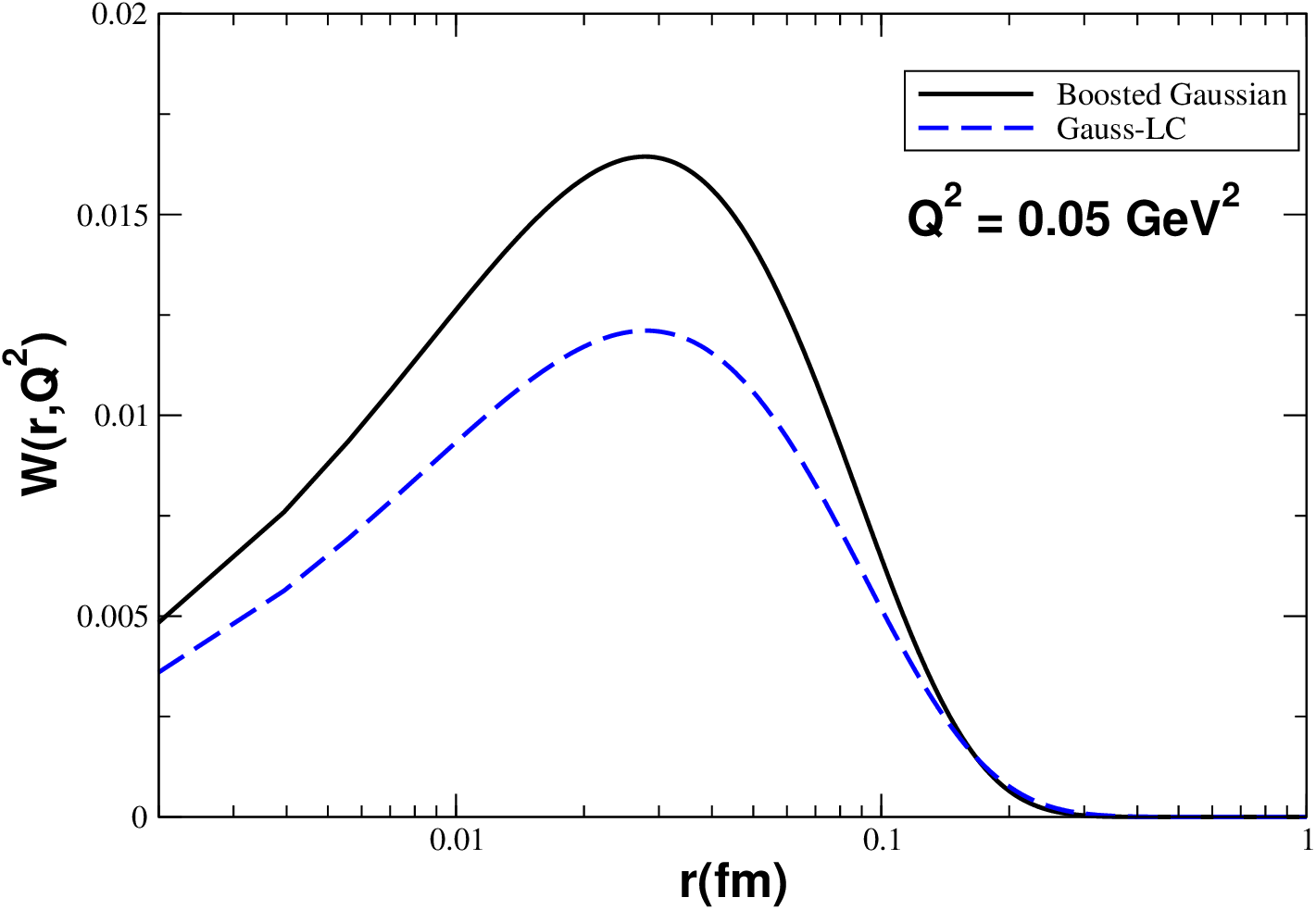} 
\caption{(Color online) Overlap function between the photon and $\Upsilon$ wave function integrated over $z$, as defined in Eq. (\ref{dabliu}),      at $Q^2 = 0.05$ GeV$^2$. }
\label{fig1}
\end{figure}

In order to estimate the total cross section we need to specify the overlap function $(\Psi_{\Upsilon}^* \Psi)$ and the non-forward scattering amplitude ${\cal{N}}(x,\rr,\rb)$. Initially let us discuss the models used for the overlap function.
In contrast to the photon wave function, which is well known in the literature (See e.g. \cite{KMW}), the description of the $\Upsilon$ wave function still is an open question. 
The simplest approach  is to assume that the vector meson is predominantly a quark-antiquark state and that the spin and polarization structure is the same as in the  photon \cite{dgkp,nnpz,sandapen,KT}. As a consequence, the overlap between the photon and the vector meson wave function, for the transversely polarized  
case, is given by (For details see Ref. \cite{KMW})
\begin{eqnarray}
(\Psi_{V}^* \Psi)_T = \hat{e}_f e \frac{N_c}{\pi z (1-z)}\left\{m_f^2K_0(\epsilon r)\phi_T(r,z) -[z^2+(1-z)^2]\epsilon K_1(\epsilon r) \partial_r \phi_T(r,z)\right\} \,\,,
\end{eqnarray}
where $ \hat{e}_f $ is the effective charge of the vector meson, $m_f$ is the quark mass, $N_c = 3$, $\epsilon^2 = z(1-z)Q^2 + m_f^2$ and $\phi_T(r,z)$ define the scalar part of the  vector meson wave function. In what follows we will consider the Boosted Gaussian and Gauss-LC models  for $\phi_T(r,z)$, which are largely used in the literature.
In the Boosted Gaussian model the function $\phi_T(r,z)$ is given by
\begin{eqnarray}
\phi_T(r,z) = N_T z(1-z) \exp\left(-\frac{m_fR^2}{8z(1-z)} - \frac{2z(1-z)r^2}{R^2} + \frac{m_f^2R^2}{2}\right) \,\,.
\end{eqnarray}
 In contrast, in the Gauss-LC model, it is given by
\begin{eqnarray}
\phi_T(r,z) = N_T [z(1-z)]^2 \exp\left(-\frac{r^2}{2R_T^2}\right)
\end{eqnarray}
The parameters $N_T$, $R$ and $R_T$ are  determined by the normalization condition of the wave function and by the decay width. In Table I  we present the value of these parameters for the $\Upsilon$ wave function. In order to analyse  the $r$-dependence of the overlap function predicted by these two models, it is useful to estimate the quantity
\begin{eqnarray}
W(r,Q^2) = 2\pi r \int \frac{dz}{4\pi} (\Psi_{\Upsilon}^* \Psi)_T\,\,.
\label{dabliu}
\end{eqnarray} 
In Fig. \ref{fig1} we present our predictions for very low $Q^2$, typical for photoproduction. We obtain that both models 
predict a peak for small values of $r$, which is directly associated to the large bottom mass. Moreover, the predicted radius dependence is similar, with the normalization of the Gauss-LC model being smaller than the Boosted Gaussian one. Such differences have direct implications in the corresponding predictions for the total cross section, as we will demonstrate in the next section.

\begin{table}[t] 
\centering
\begin{tabular}{cccccc} 
\hline 
Model & $M_{\Upsilon}/\mbox{GeV}$ & $m_{f}/\mbox{GeV}$ & $N_{T}$ & $R_T^{2}/\mbox{GeV}^{-2}$ & $R^{2}/\mbox{GeV}^{-2}$ \\ 
\hline
\hline
Gauss-LC & 9.460 & 4.2 & 0.76  & 1.91 & -- \\
Boosted Gaussian & 9.460 & 4.2 & 0.481 & -- & 0.57 \\
\hline
\label{tab1}
\end{tabular}
\caption{Parameters of the Gauss-LC and Boosted Gaussian models for the $\Upsilon$ wave function.} 
\end{table}

The non-forward scattering amplitude ${\cal{N}}(x,\rr,\rb)$   contains all
information about the target and the strong interaction physics.
In the last years, several groups have constructed phenomenological models which satisfy the asymptotic behaviour of 
the Color Glass Condensate (CGC)  formalism \cite{CGC,BAL,kov}.
In what follows we will use the 
bCGC model proposed in Ref. \cite{KMW}, which improves the Iancu - Itakura - Munier (IIM) model 
 \cite{iim} with  the inclusion of   the impact parameter dependence in the dipole - proton scattering amplitude.   Following \cite{KMW} we have:
\begin{eqnarray}
\mathcal{N}_p(x,\rr,{\rb}) =   
\left\{ \begin{array}{ll} 
{\mathcal N}_0\, \left(\frac{ r \, Q_{s,p}}{2}\right)^{2\left(\gamma_s + 
\frac{\ln (2/r Q_{s,p})}{\kappa \,\lambda \,Y}\right)}  & \mbox{$r Q_{s,p} \le 2$} \\
 1 - \exp \left[-A\,\ln^2\,(B \, r \, Q_{s,p})\right]   & \mbox{$r Q_{s,p}  > 2$} 
\end{array} \right.
\label{eq:bcgc}
\end{eqnarray}
with  $Y=\ln(1/x)$ and $\kappa = \chi''(\gamma_s)/\chi'(\gamma_s)$, where $\chi$ is the 
LO BFKL characteristic function \cite{bfkl}.  The coefficients $A$ and $B$  
are determined uniquely from the condition that $\mathcal{N}_p(x,\rr,\rb)$, and its derivative 
with respect to $rQ_s$, are continuous at $rQ_s=2$. 
In this model, the proton saturation scale $Q_{s,p}$ depends on the impact parameter:
\begin{equation} 
  Q_{s,p}\equiv Q_{s,p}(x,{\rb})=\left(\frac{x_0}{x}\right)^{\frac{\lambda}{2}}\;
\left[\exp\left(-\frac{{b}^2}{2B_{\rm CGC}}\right)\right]^{\frac{1}{2\gamma_s}}.
\label{newqs}
\end{equation}
The parameter $B_{\rm CGC}$  was  adjusted to give a good 
description of the $t$-dependence of exclusive $J/\psi$ photoproduction.  
The factors $\mathcal{N}_0$, $x_0$,  $\lambda$ and  $\gamma_s$  were  taken  to be free. Recently the parameters of this model have been updated in Ref. \cite{amir} (considering the 
recently released high precision combined HERA data), being given by  $\gamma_s = 0.6599$, $B_{CGC} = 5.5$ GeV$^{-2}$,
$\mathcal{N}_0 = 0.3358$, $x_0 = 0.00105 \times 10^{-5}$ and $\lambda = 0.2063$. As demonstrate in Ref. \cite{armesto_amir}, this phenomenological dipole  describes quite well the HERA data for the exclusive $\rho$ and $J/\Psi$ production. 
For comparison, in what follows we will also  use the GBW model \cite{GBW}, which  assumes that 
${\cal{N}}_p(x,\rr,\rb) = {\cal{N}}_p(x,\rr) S(\rb)$ with the forward scattering amplitude being given by ${\cal{N}}_p(x,\rr)=1-e^{-\rr^2Q_{s,p}^2(Y)/4}$ and $Q_{s,p}^2(Y)=\left(x_0/x\right)^{\lambda}$, with the parameters $x_0$ and $\lambda$ determined by the fit to the HERA data available in 1999.
The parameters of the GBW model have been updated in Ref. \cite{kozlov} considering the ZEUS data available in 2007.
In what follows we will use these two sets of parameters in our calculations, with the resulting predictions being denoted GBW and GBW-KSX, respectively. 
It is important to emphasize that the GBW model is a model for the  forward dipole-target amplitude ${\cal{N}}_p(x,\rr)$, which does not allow us 
to calculate the $t$-dependence of the differential cross section. Therefore, in the GBW case, we should  estimate the total cross section using  
Eq. (\ref{dsdt_approx}).

\begin{figure}
\begin{tabular}{cc}
\includegraphics[scale=0.35]{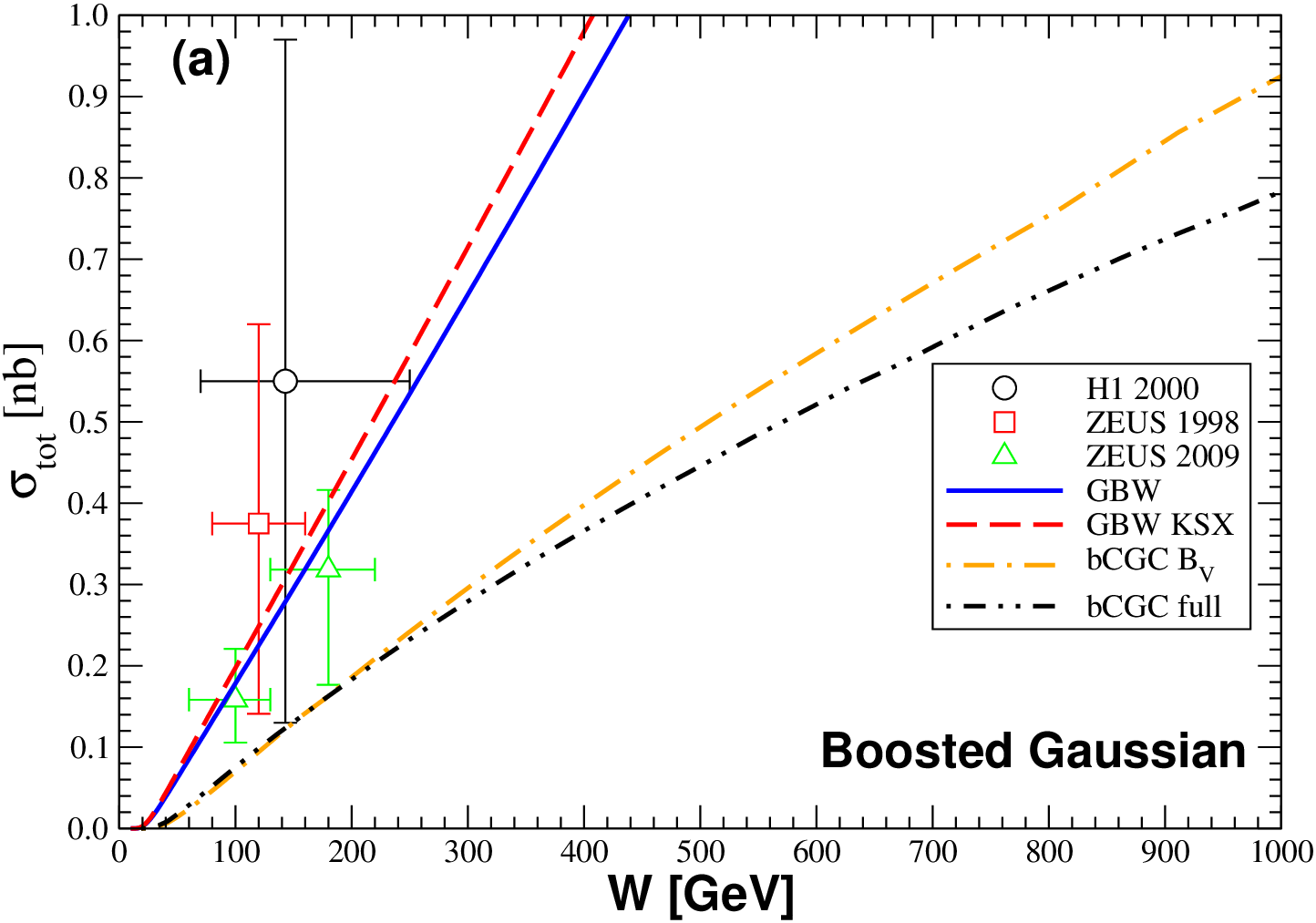} & 
\includegraphics[scale=0.35]{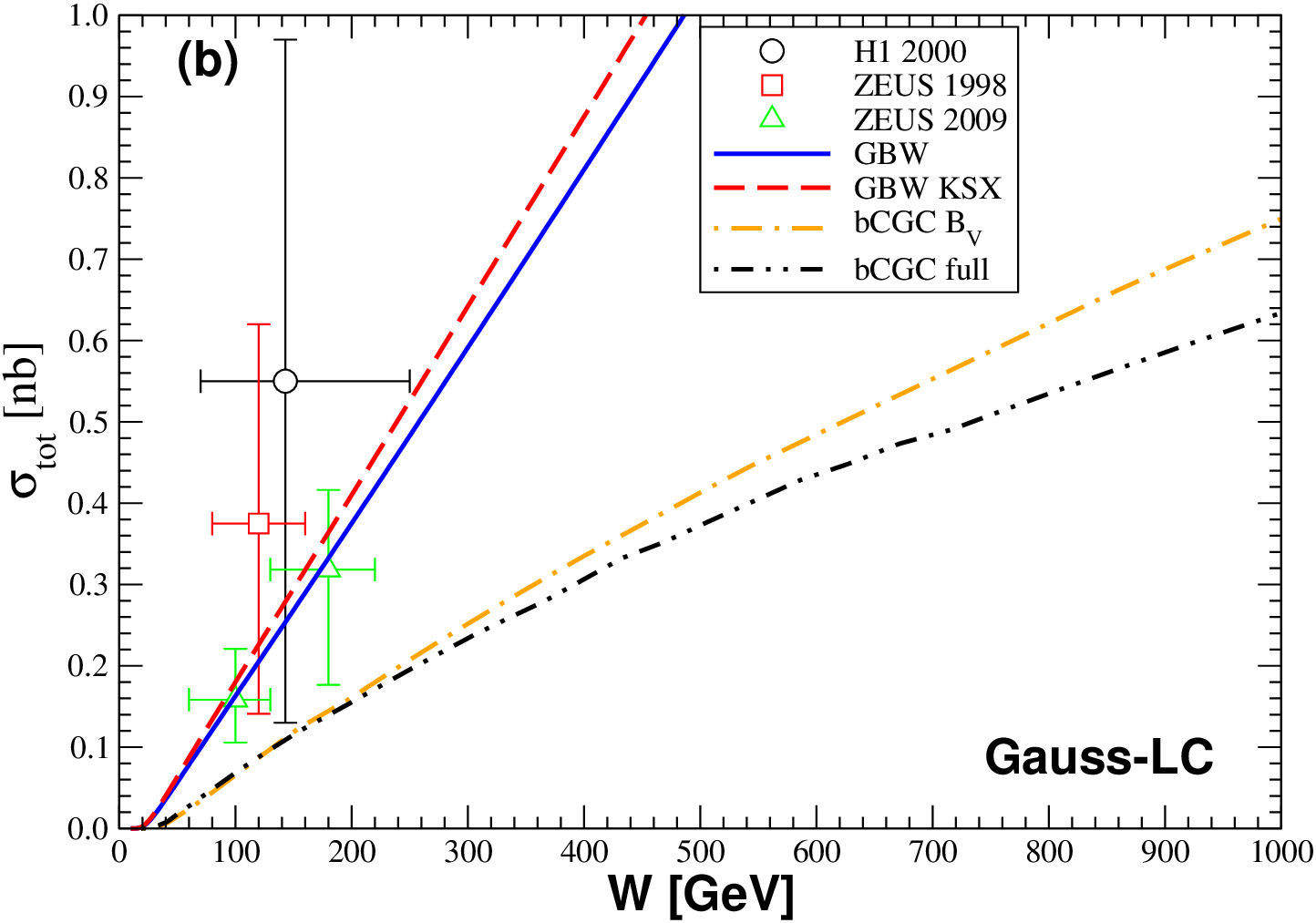}
\end{tabular}
\caption{(Color online) Exclusive $\Upsilon$ photoproduction in  $\gamma p$  
 collisions. Data from HERA \cite{hera_data_ups}. }
\label{fig2}
\end{figure}

\begin{figure}
\begin{tabular}{cc}
\includegraphics[scale=0.35]{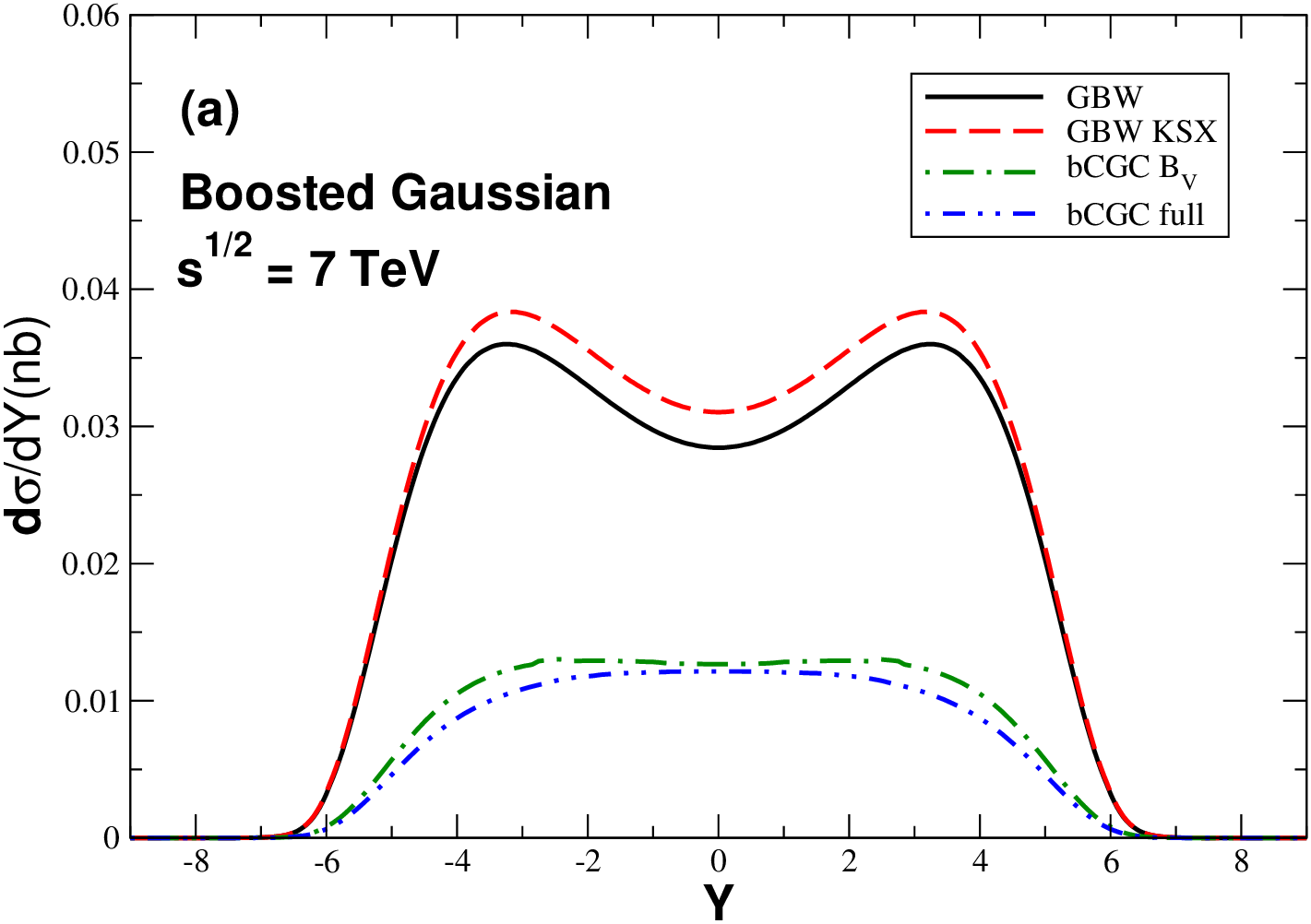} & 
\includegraphics[scale=0.35]{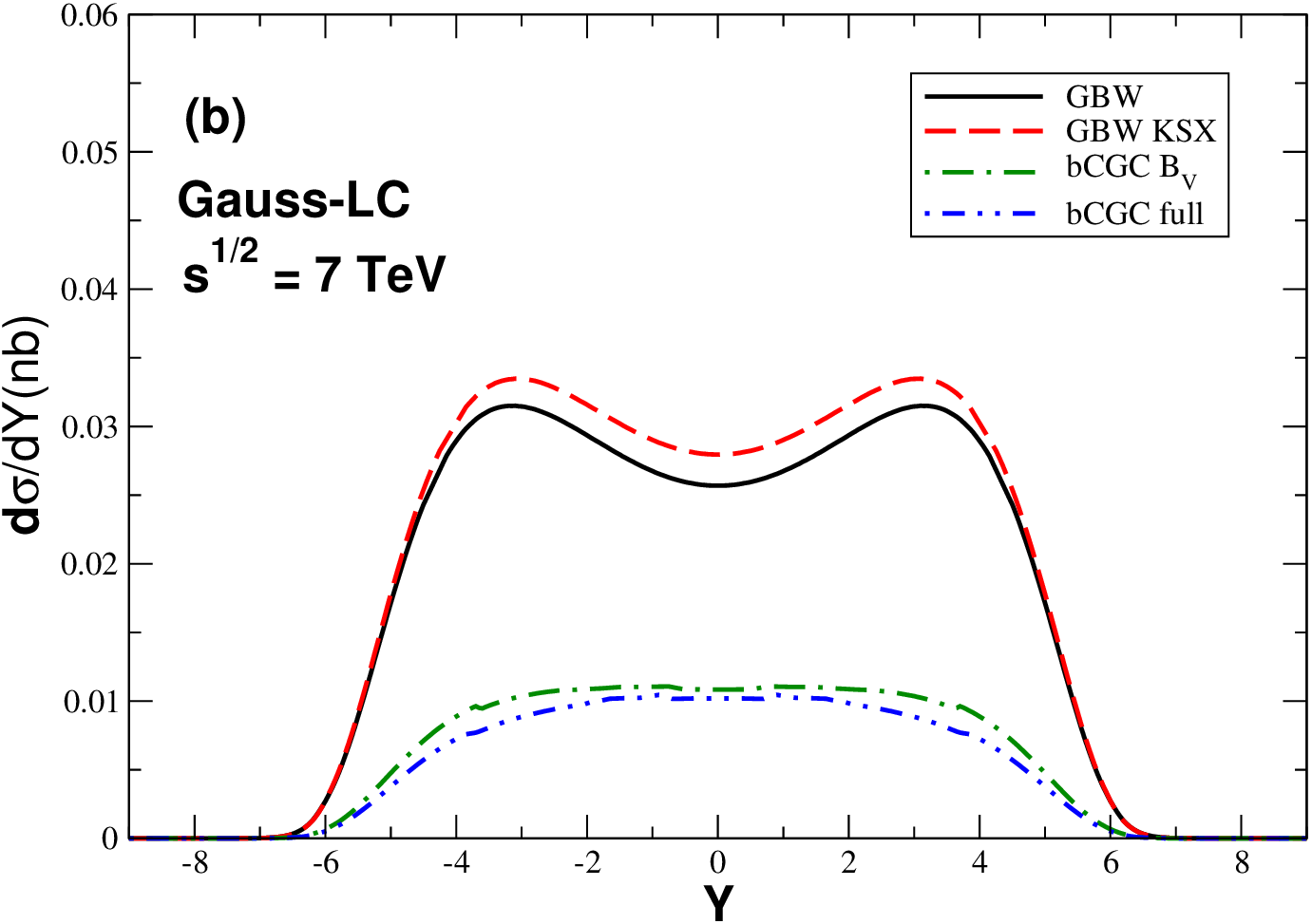}
\end{tabular}
\caption{(Color online) Exclusive $\Upsilon$ photoproduction in $pp$ collisions at $\sqrt{s} = 7$ TeV. }
\label{fig3}
\end{figure}

\begin{figure}
\begin{tabular}{cc}
\includegraphics[scale=0.35]{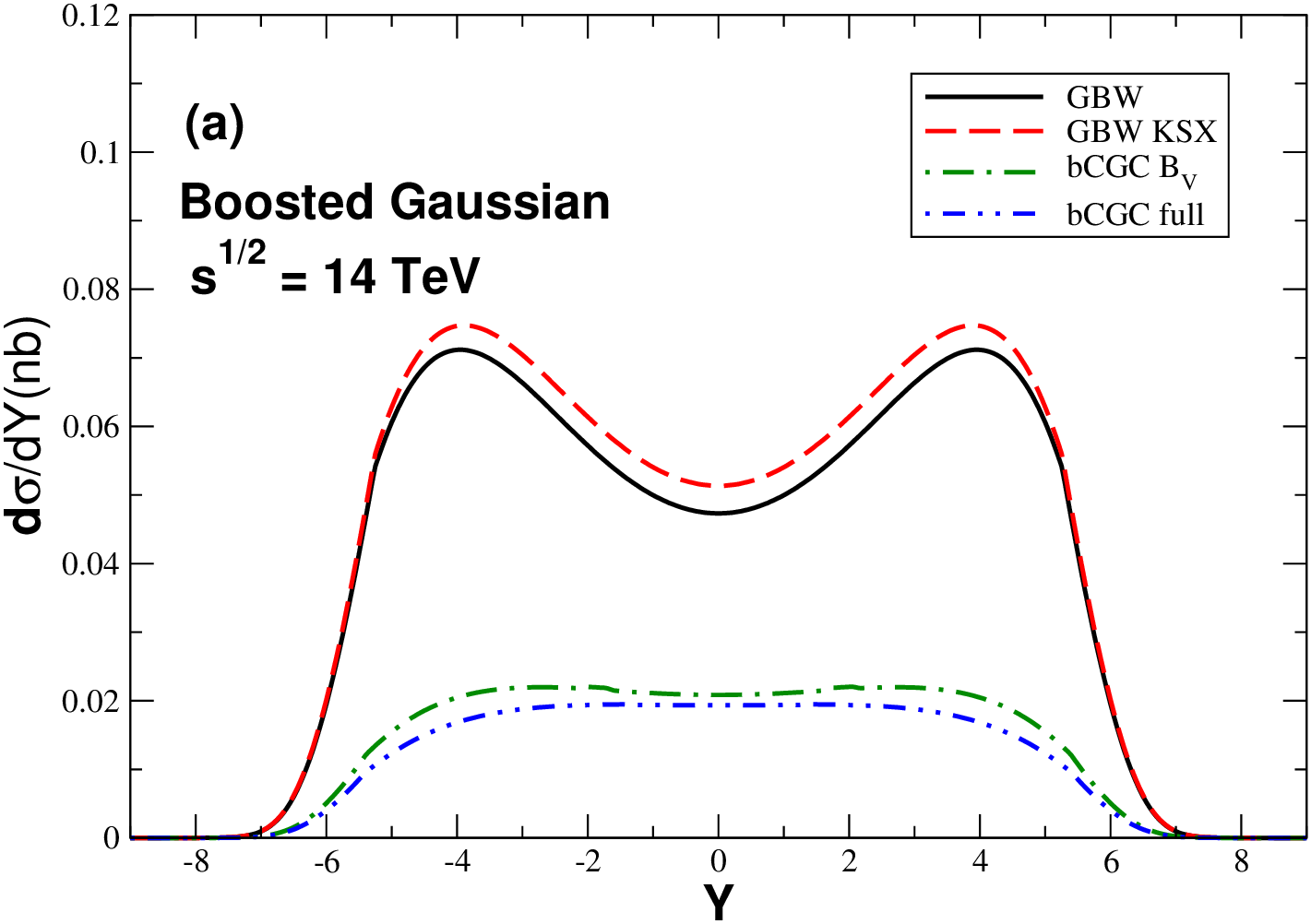} & 
\includegraphics[scale=0.35]{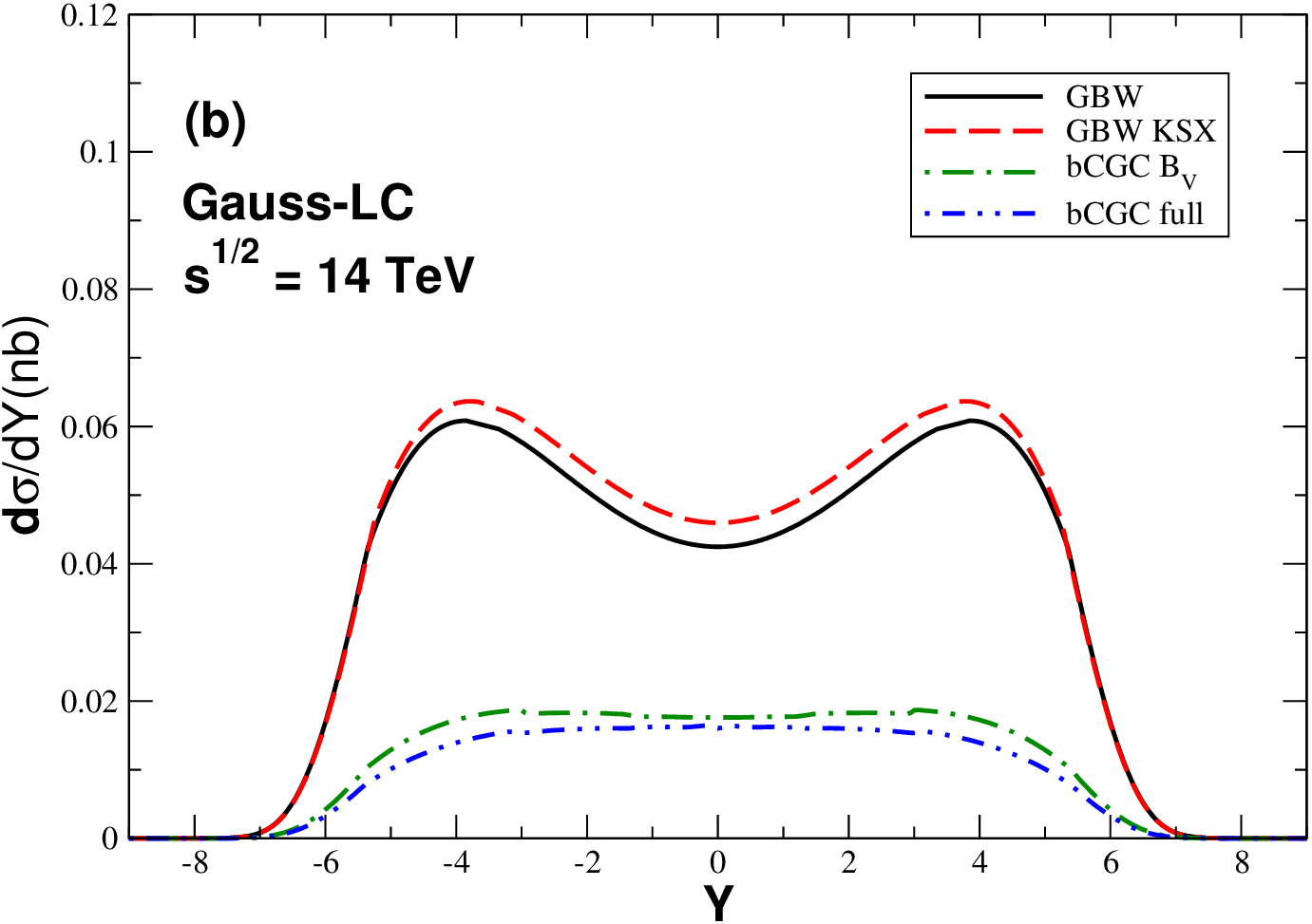}
\end{tabular}
\caption{(Color online) Exclusive $\Upsilon$ photoproduction in $pp$ collisions at $\sqrt{s} = 14$ TeV. }
\label{fig4}
\end{figure}

\begin{figure}
\begin{tabular}{cc}
\includegraphics[scale=0.35]{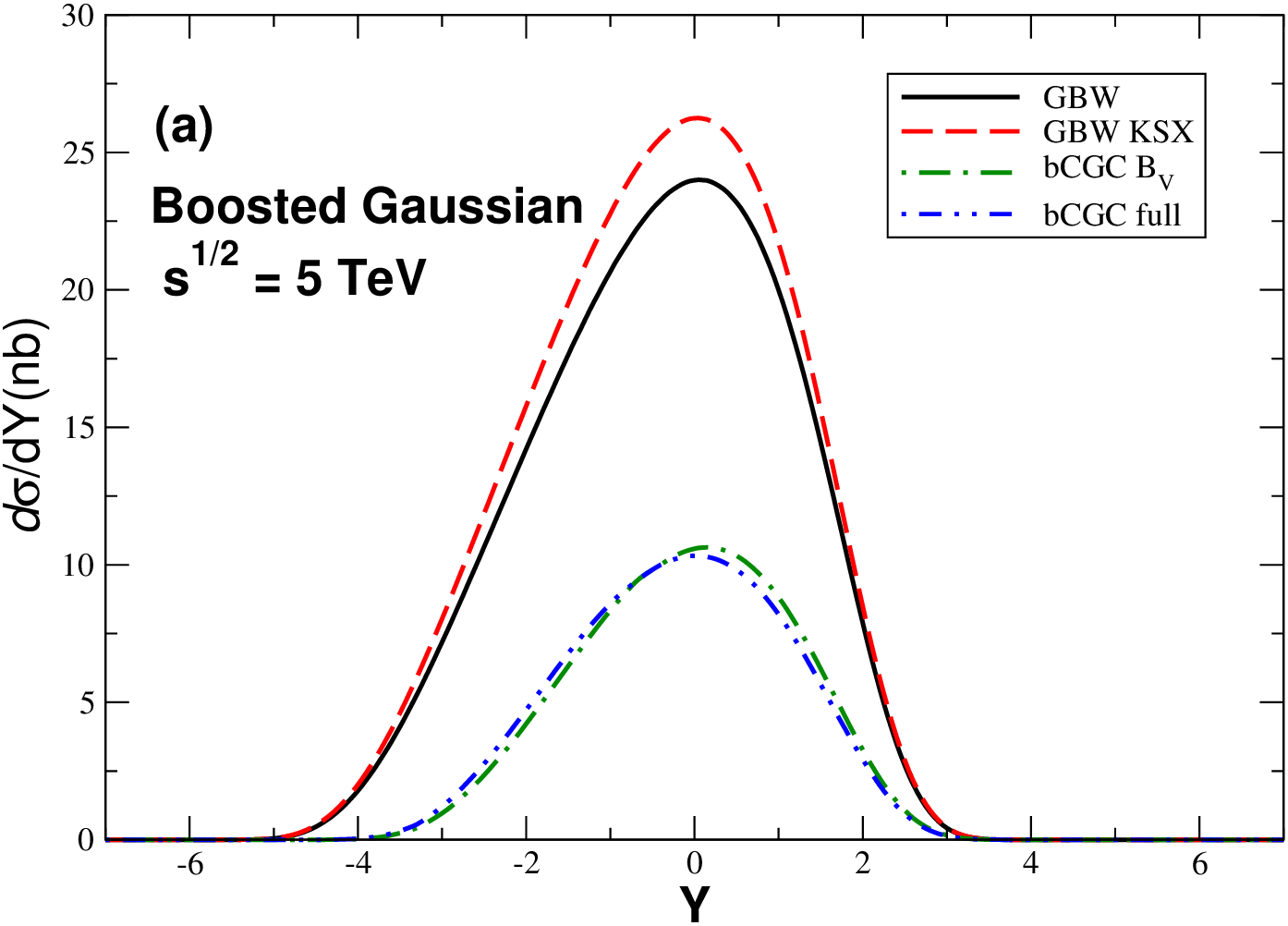} & 
\includegraphics[scale=0.35]{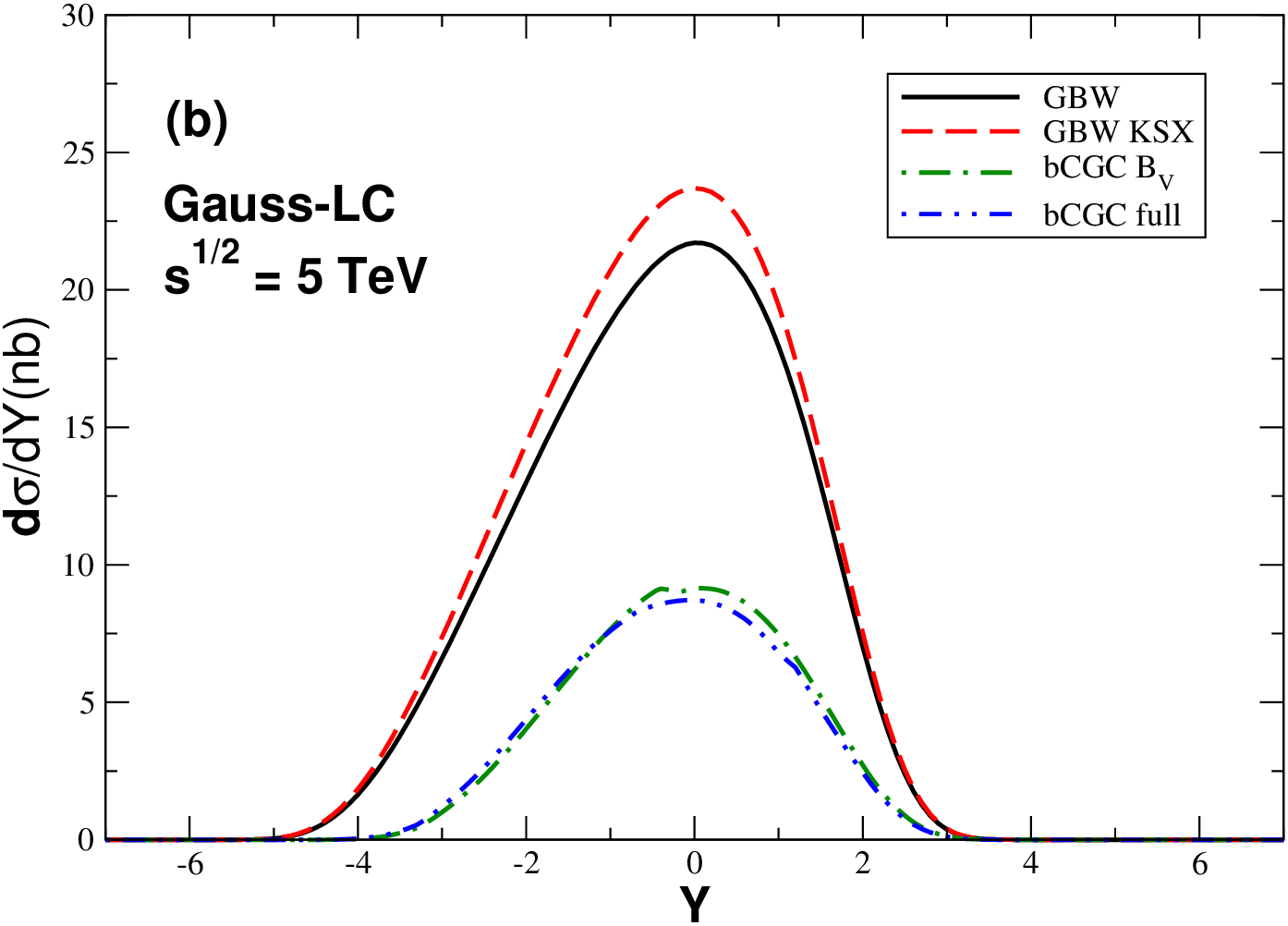}
\end{tabular}
\caption{(Color online) Exclusive $\Upsilon$ photoproduction in $pPb$ collisions at $\sqrt{s} = 5$ TeV.}
\label{fig5}
\end{figure}

\section{Results}
\label{resultados}

In what follows we will present our predictions for the exclusive $\Upsilon$ photoproduction in photon - proton and proton - proton/nucleus collisions. In order to estimate the cross section we need to specify the exponent $\lambda_e$ which determines $R_g$ and 
$\beta$. As demonstrated in Ref. \cite{harland} the estimate obtained using this approximation for $R_g$ is strongly dependent on the parton distribution used in the calculation. However, this dependence is smaller at large hard scales and small values of $x$ (large energies), which is the case of the $\Upsilon$ production at LHC. Another important aspect is that the incorporation of the skewness correction at small-$x$ in the dipole models still is an open question which deserve more detailed studies (See  e.g. Ref. \cite{amir}).     Following Ref. \cite{amir} we will disregard that $\lambda_e$ can be scale dependent, which is good approximation at large hard scales, and we will assume that it is equal to the exponent $\lambda$ that determines the energy dependence of the saturation scale.  Consequently, our estimate for the factor $R_g$ should be considered a phenomenological estimate. For a more detailed discussion about the subject see Ref. \cite{harland}. 
Moreover, in our calculations of the exclusive $\Upsilon$ photoproduction in $pp$ and $pPb$ collisions
 we will disregard   soft interactions which lead to an extra production of particles that destroy the rapidity gap in the final state. The inclusion of these additional absorption effects can be parametrized in terms of a multiplicative factor denoted rapidity gap survival probability, $S^2$, which corresponds to the probability of the scattered proton not to dissociate due to the secondary interactions. 
In Ref. \cite{Schafer}  the authors have estimated $S^2$ and obtained that in $pp/p\bar{p}$ collisions it is $ \sim 0.8 - 0.9$,  depending  on the  rapidity of the vector meson (See also Refs. \cite{Guzey,Martin}).

In Figs. \ref{fig2} (a) and (b) we present our predictions for the energy dependence of  exclusive $\Upsilon$ photoproduction in $\gamma p$ collisions considering the Boosted Gaussian and Gauss-LC models for the $\Upsilon$ wave function, respectively.  We denote by bCGC full the predictions obtained using Eq. (\ref{sctotal_intt}), i.e. taking into account the $t$-dependence of the differential cross section. We also present the prediction obtained using the exponential approximation as given by the  Eq. (\ref{dsdt_approx}), denoted bCGC $B_V$ hereafter. For comparison we also present the GBW and GBW-KSX predictions. We obtain that the Boosted Gaussian predictions are always larger than the Gauss-LC one, as expected from Fig. \ref{fig1}. The scarce HERA data are reasonably described by the two versions of the GBW model, with the  bCGC one  underestimating the data, in agreement with previous results \cite{motyka_watt,cox_forshaw} obtained using other models for the dipole - proton scattering amplitude. A possible explanation for the difference between the GBW and bCGC predictions is the distinct behaviour of ${\cal{N}}_p$ at  small values of $\rr$ predicted  by these models.  As the total cross section for the $\Upsilon$ production is dominated by very small values of the radius, it is probing the linear behaviour of   ${\cal{N}}_p$. While the GBW model predicts that ${\cal{N}}_p \propto \rr^2$, the bCGC one predicts ${\cal{N}}_p \propto \rr^{2\gamma_{eff}}$, with $\gamma_{eff} \le 1$.
Another important aspect to be emphasized is that our results demonstrate that the approximation of the $t$-dependence by an exponential form is a reasonable 
approximation for $W \le 200$ GeV, but overestimates the cross section for larger values of the $\gamma p$ center-of-mass energy.

In Figs. \ref{fig3} and \ref{fig4} we present our predictions for the rapidity distribution of exclusive $\Upsilon$ photoproduction  in $pp$ collisions  at $\sqrt{s}=7$ TeV and  $\sqrt{s}=14$ TeV, respectively. The Boosted Gaussian and Gauss-LC predictions are presented in the panels (a) and (b), respectively. We obtain that the differences between the predictions observed in Fig. \ref{fig2} are also present in the rapidity distribution, with the GBW-KSX (bCGC full) prediction being an upper (lower) bound for the predictions at $Y = 0$. For $\sqrt{s}=7$ TeV we obtain that he  bCGC full and $B_V$ predictions are almost identical at central rapidities and differ by $\approx 10$ \% for $Y = 4$. The GBW-KSX and GBW predictions differ by $\approx 10$ \% in the $|Y| \le 4$ range. In contrast, the GBW and bCGC predictions differ by a factor 2.7 (3.5) at $Y = 0 \, (4)$, which is directly associated to a distinct energy dependence of the $\gamma p$ cross section observed in Fig. \ref{fig2}.
As also expected from Fig. \ref{fig2}, the Boosted Gaussian predictions are larger than the Gauss-LC one, with the difference being of $\approx 12$ \% at $Y = 0$. For $\sqrt{s}=14$ TeV we obtain similar results, with the main difference being the larger values for the rapidity distribution. In comparison to the results presented in Ref. \cite{Martin}, 
which predict the rapidity distribution for $\Upsilon$ production at LHC considering leading order (LO) and next-to-leading order (NLO) corrections to the  exclusive photon - hadron cross section, our GBW predictions are very similar to those associated to the LO fit, while our bCGC predictions are similar to the NLO one.

\begin{table}[t] 
\centering
\begin{tabular}{ccccc} 
\hline 
\hline
       &  GBW  & GBW KSX & bCGC $B_{V}$ & bCGC full \\ 
\hline
\hline
      & &{\bf Gauss-LC} & & \\  
\hline 
$pp$ ($\sqrt{s} = $ 7 TeV) & 298.0 pb & 318.0 pb  & 104.0 pb & 91.0 pb  \\
$pp$ ($\sqrt{s} = $ 8 TeV) & 344.0 pb & 366.0 pb & 118.0 pb & 103.0 pb  \\
$pp$ ($\sqrt{s} = $ 14 TeV) & 607.0 pb & 638.0 pb & 196.0 pb  & 167.0 pb  \\
$p\mbox{Pb}$ ($\sqrt{s} = $ 5 TeV) & 87.1 nb & 95.4 nb & 32.222 nb & 31.3 nb \\
\hline
\hline
     & & {\bf Boosted Gaussian} & & \\
\hline 
$pp$ ($\sqrt{s} = $ 7 TeV) & 340.0 pb & 363.0 pb & 123.0 pb & 110.0 pb \\
$pp$ ($\sqrt{s} = $ 8 TeV) & 393.0 pb & 419.0 pb & 140.0 pb & 124.0 pb \\
$pp$ ($\sqrt{s} = $ 14 TeV) & 699.0 pb & 740.0 pb & 233.0 pb  & 201.0 pb \\
$p\mbox{Pb}$ ($\sqrt{s} = $ 5 TeV) & 96.1 nb & 105.5 nb & 36.374 nb & 36.1 nb \\
\hline
\end{tabular}
\caption{Total cross sections for the exclusive $\Upsilon$ photoproduction in $pp$ collisions at $\sqrt{s} = 7, \,8$ and 14 TeV and $pPb$ collisions at $\sqrt{s} = 5$ TeV considering the Gauss-LC and Boosted Gaussian models for the vector meson wave function.} 
\label{tab2}
\end{table}

In Fig. \ref{fig5} we present our predictions for the rapidity distribution for the exclusive $\Upsilon$ photoproduction   in $pPb$ collisions  at $\sqrt{s}=5$ TeV. As expected, the rapidity distribution is asymmetric with respect to $Y=0$, being dominated by $\gamma p$ interactions, due to the $Z^2$ enhancement present in the nuclear photon spectrum. We observe that the predictions differ by a factor 2.6 at $Y= 0$. Finally, in Table \ref{tab2} we present our predictions for the total cross section for the exclusive $\Upsilon$ photoproduction   in $pp$ and $pPb$  collisions at  LHC energies. In particular, in Table \ref{tab3} we present our predictions for the $\Upsilon$ photoproduction in the LHCb kinematical range ($2 \leq Y \leq 4.5$). As expected from our analysis of the rapidity distributions, the predictions for the total cross sections are largely distinct.

\section{Summary}
\label{sumario}
Recent experimental results have demonstrated that the study of the QCD dynamics using photon induced interactions in hadronic collisions is feasible and that it is possible to probe several aspects of the hadronic physics. In particular, $\gamma \gamma$ and $\gamma h$ interactions at LHC are probing a kinematical range unexplored by previous colliders. The results for  exclusive $J/\Psi$ photoproduction are allowing to extend the studies performed at HERA and to obtain more informations about the high energy 
behaviour of the QCD dynamics as well as about the vector meson wave function. A similar expectation exists for the exclusive $\Upsilon$ photoproduction in  $pp$ and $pPb$ collisions. Although this process has been studied before,  different assumptions for the meson wave function and QCD dynamics, as well as for the free parameters, have been considered in these analysis. Our goal in this paper was, using the color dipole formalism, to estimate the theoretical uncertainty associated to the description of the QCD dynamics. We have assumed two distinct  models for the $\Upsilon$ wave function and considered three  models for the dipole - proton scattering amplitude. Moreover, we have compared the results obtained considering the $t$-dependence of the differential cross section with the  exponential approximation and verified that their predictions at large energies are distinct. 
We demonstrated that although these models satisfactorily 
describe the HERA data, their predictions are very distinct for the exclusive $\Upsilon$ photoproduction  in $pp/pPb$ collisions. Furthermore, we present our predictions for the LHCb kinematical range. Our main conclusion is that  future measurements can be useful to constrain the magnitude of the nonlinear effects in the QCD dynamics 
as well as  models for the vector meson wave function.

\begin{table}[t] 
\centering
\begin{tabular}{ccccc} 
\hline 
\hline
       &  GBW  & GBW KSX & bCGC $B_{V}$ & bCGC full \\ 
\hline
\hline
      & &{\bf Gauss-LC} & & \\  
\hline 
$pp$ ($\sqrt{s} = $ 7 TeV) & 75.0 pb & 79.0 pb & 24.0 pb & 21.0 pb  \\
$pp$ ($\sqrt{s} = $ 8 TeV) & 86.0 pb & 91.0 pb & 28.0 pb & 23.0 pb  \\
$pp$ ($\sqrt{s} = $ 14 TeV) & 144.0 pb & 151.0 pb & 45.0 pb & 37.0 pb \\
$p\mbox{Pb}$ ($\sqrt{s} = $ 5 TeV) & 2.9 nb & 3.1 nb & 1.1 nb & 0.96 nb \\
\hline
\hline
     & & {\bf Boosted Gaussian} & & \\
\hline 
$pp$ ($\sqrt{s} = $ 7 TeV) & 86.0 pb & 91.0 pb & 29.0 pb & 25.0 pb \\
$pp$ ($\sqrt{s} = $ 8 TeV) & 98.0 pb & 104.0 pb & 33.0 pb & 28.0 pb \\
$pp$ ($\sqrt{s} = $ 14 TeV) & 166.0 pb & 176.0 pb & 53.0 pb & 45.0 pb \\
$p\mbox{Pb}$ ($\sqrt{s} = $ 5 TeV) & 3.3 nb & 3.5 nb & 1.3 nb & 1.2 nb \\
\hline
\end{tabular}
\caption{Total cross sections for the exclusive $\Upsilon$ photoproduction in the LHCb kinematical range  ($2 \leq Y \leq 4.5$).} 
\label{tab3}
\end{table}


\section*{Acknowledgements}

VPG thanks Gustavo Gil da Silveira and Murilo Rangel for useful discussions. This work was partially financed by the Brazilian funding agencies CAPES, CNPq,  FAPESP and FAPERGS.



\end{document}